\documentclass[fleqn,twoside]{article}
\usepackage{espcrc2}

\usepackage{graphicx}
\newcommand{\Xmaxi}{X_{\max,i}}
\newcommand{\Xmax}{X_{\max}}
\newcommand{\Smax}{S_{\max}}
\newcommand{\Smaxi}{S_{\max,i}}
\newcommand{\nexus}{\textsc{neXus~3.97}}
\newcommand{\sibyll}{\textsc{sibyll~2.1}}
\newcommand{\qgsjet}{\textsc{qgsjet01}}
\newcommand{\conex}{\textsc{conex}}
\newcommand{\seneca}{\textsc{seneca}}
\newcommand{\corsika}{\textsc{corsika}}
\title{Hybrid Simulation of Cosmic Ray Air Showers}

\author{H.J. Drescher \address[MCSD]{ 
\small Institut f\"ur Theoretische Physik,
Johann Wolfgang Goethe-Universit\"at,
Robert-Mayer-Str. 8-10,
60325~Frankfurt am Main, Germany}}
       
\begin{document}

\begin{abstract}
Air shower simulations are essential for interpreting
data from cosmic ray experiments. 
At highest energies though, a
microscopic treatment of a whole shower is not possible any more,
since it would require a huge amount of CPU-time. 
We review hybrid approaches of air shower simulation which try to
overcome this problem without giving rise to artificial fluctuations
as generated by the thinning algorithm.

\vspace{1pc}
\end{abstract}

\maketitle

\section{Introduction}

Ultra-high energy cosmic rays (UHECR) are currently of great
interest. One still ignores the nature of these particles, 
acceleration mechanisms and possible sources. Direct measurements at
these energies are impossible due to a very low flux. Upon entering
the atmosphere, cosmic rays induce extensive air showers (EAS),
cascades of particles produced by collisions of the primary and
subsequent secondaries with air nuclei. Experiments measure these
showers and try to deduce properties of the primary particle from
properties of the shower. One can therefore 
consider the atmosphere as a huge detector in which a primary
cosmic ray is absorbed and observed.

There are basically two types of experimental set-ups. Ground arrays
(i.e. AGASA, Auger) measure the density of charged particles on the
ground. The lateral distribution function (LDF) is studied to deduce
the energy. One can estimate the arrival direction from the different
trigger-times of the detectors as the shower front passes through the
array. 
The other kind are optical detectors. They 
collect the light emitted by fluorescence of
nitrogen molecules, which are excited by the flux of charged
particles going through the atmosphere. They measure therefore
directly the longitudinal profile of a shower. Both types of
experiments have advantages and drawbacks. Ground arrays have a high
duty cycle and are not very sensitive to details of the atmosphere.
Fluorescence experiments depend somewhat less on models, but are
influenced by atmospheric fluctuations and have a duty cycle of
typically 10\%. 

For both set-ups, reconstruction of primary
properties depends on how good one understands the interactions in the
atmosphere. 
Air shower simulations are therefore crucial for cosmic ray physics as
they are needed for interpretation of the data. 
A major difficulty is that no experimental data from accelerators
is available at these energies.
 LHC is still three orders of magnitudes
(lab. system) lower than the currently highest energies measured. 
Another problem is that air shower development is dominated by forward
scattering, since most energy is carried by the leading
particles, but accelerators measure mostly at mid-rapidity. Also, the targets
are light nuclei, commonly less well studied. 
Therefore, physics has to be extrapolated into unknown regions (energy and
phase-space) which makes interpretation of data less precise. 
On the other hand, one can argue that cosmic rays provide a unique
opportunity to study physics at high energies that may never be
reached by accelerator experiments. 

Air showers develop rapidly in the atmosphere. A rule of thumb is that
the number of charged particles at the maximum of the longitudinal
profile is about 60\% of the energy measured in GeV; a $10^{11}$ GeV
shower has about 60 billion particles. Hence,
microscopic simulations seem quite impossible. E.g., a $10^{10}$~GeV shower
would take about a year to compute on current CPUs. 
The thinning algorithm \cite{thin} reduces the CPU time but introduces artificial
fluctuations. 
Hybrid simulations  \cite{Ded68} try to solve this problem by
following the high energy part of an air shower in detail, and using
efficient approximations below a given threshold.

\section{The thinning problem in microscopic treatment of air shower
  simulations} 

Current air shower simulations use the thinning algorithm in order to
limit the computation time. The original idea is to follow only one
particle of all secondaries below a given
threshold $E_{th}=f_{th} E_0$, chosen with the probability $p_i=E_i/E_{\rm
  tot}$. Neglected particles are compensated by attributing a higher
weight $w_i=1/p_i$ to the chosen particle. 
In principle, the probability of some particle below the threshold to be
followed can be an arbitrary function, but the weight attributed has to
be $1/p_i$. Statistical thinning means $p_i=E_i/E_{\rm tot}$ (as above,
but the number of followed particles is not fixed to one). 
Further refinements are possible by imposing a weight limit on the
algorithm, or not to thin beyond a given distance from the
shower axis. The disadvantage of thinning is that it introduces
artificial fluctuations, and one has to be careful to control these. 

The hybrid approach tries to overcome these problems by computing the
sub-showers of particles efficiently instead of  discarding these.

\begin{figure}[tbh]
\includegraphics[width=\columnwidth]{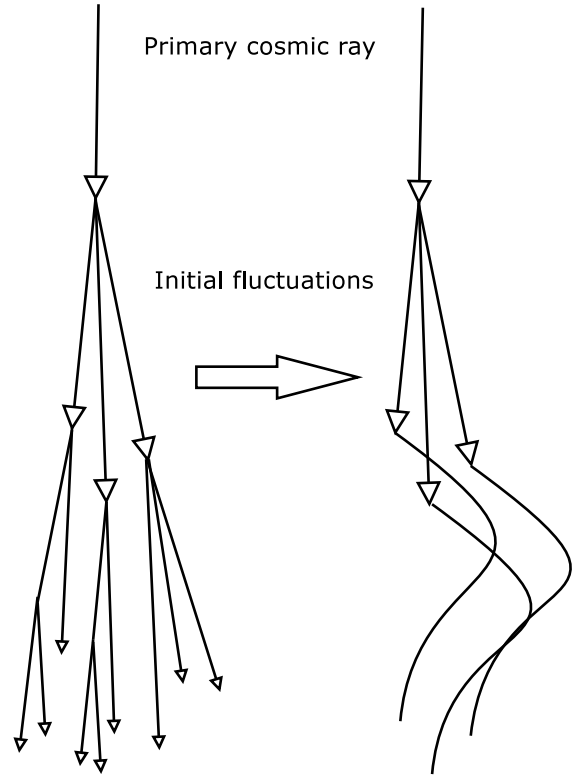}
\caption{Principle of the hybrid approach: compute sub-showers with
  efficient algorithms.}
\label{fig:fig1}
\end{figure}

\section{Hybrid approaches}

As mentioned, hybrid methods replace sub-threshold particles with 
the sub-showers induced by them. We are going to discuss two
different approaches, where these sub-showers are obtained  
from a shower library or the solution of cascade equations. 
  They differ also in the fact, that the former
replaces a fluctuating shower whereas the latter solves for mean
sub-showers. 

\subsection{Bartol approach to hybrid simulations}

An obvious choice is to implement a data base of pre-simulated
showers. During the simulation, all sub-showers
below a given energy threshold are replaced with a sample from
this data base. 
The Bartol \cite{Alvarez-Muniz:2002ne} approach consists of a shower
library of pion-initiated 
showers for different injection depths, energies and zenith angles.
Nucleons are followed by Monte-Carlo method, whereas kaons are treated
to behave as pions with respect to interaction. 
The resulting longitudinal profiles are then fitted to a Gaisser-Hillas
function, and the parameters are recorded. In addition, the number of muons
at ground level for different energy thresholds is stored. 
 The library itself is built with a boot-strap method: Starting from
 low energy, the pre-simulated showers are used for the computation of
 higher energies. 
During the simulation of a given shower, each particle falling below
the threshold $E_{\rm th}=0.01 E_0$ is replaced with a sub-shower
sampled from the data base. Note that these sub-showers are not average
profiles but account fully for fluctuations even below the
hybrid threshold. 

\subsection{Hybrid approach with cascade equations}

Cascade equations are one-dimensional transport equations which are solved
numerically. The solution is a mean shower profile without any fluctuations. 
Given a number of particles $h_{n}(E,X)dE$, of type
$n$, at given altitude $X$ and energy $E$, the probability for
interaction and decay is
\begin{equation}
\frac{\partial h_{n}(E,X)}{\partial X} =  -h_{n}(E,X) \left(
  \frac{1}{\lambda _{n}(E)}+\frac{d_n}{E \rho(X)} \right)
\end{equation}
with $\lambda$ being the mean free path of the particle as a function
of the energy, $\rho$ being the density of the air, and
$d_n=m_n/(c\tau_n )$ being the decay constant ($E/d_n$ is the decay
length in the lab. system).

Accounting for particles produced at higher energies
gives rise to the following system of hadronic cascade
equations \cite{bossard}:
\begin{eqnarray}
\frac{\partial h_{n}(E,X)}{\partial X} =  -h_{n}(E,X)\left[
  \frac{1}{\lambda _{n}(E)}+\frac{d_{n}}{E\rho(X)}\right] \label{for:hce} \\ 
  +\sum _{m}\int_{E}^{E_{\mathrm{max}}}h_{m}(E',X)
 \left[
  \frac{W_{mn}(E',E)}{\lambda _{m}(E')} \right. \nonumber  \\
 \left. + \frac{d_{m}D_{mn}(E',E)}{E'\rho(X)}\right] dE'\nonumber ~~. 
\end{eqnarray}
The functions \( W_{mn}(E',E) \) are the
energy-spectra \( \frac{dN}{dE} \) of secondary particles of
type \( n \) in a collision of hadron \( m \) with air-molecules.
\( D_{mn}(E',E) \) are the corresponding decay-functions. Equation
(\ref{for:hce}) is a typical transport equation with a source term.
The first term with the minus-sign accounts for particles disappearing
by collisions or decays, whereas the source term accounts for production
of secondary particles by collisions or decays of particles at higher
energies. 

The initial condition for the primary cosmic ray is given by 
\begin{equation}
\label{for:initial} h_{n}(E,X=X_{m})=\delta _{nm}\delta (E-E_{m}) .
\end{equation}
When using the hybrid approach, the initial condition is a
superposition of all particles $E<E_{th}$ produced above the
threshold.

\subsection{Low energy source functions}
At low energies, the three-dimensional spread of particles becomes
important. For the computation of the longitudinal profile only, 
 one can apply corrections to the one-dimensional
cascade equations in order to account for neglecting the lateral
expansion. This is done in the \conex ~model \cite{Pierog}.

For the computation of lateral distribution functions one can switch
back to the Monte-Carlo scheme. Particles are generated from the
so-called source function,

\begin{equation}
\frac{\partial h^{\rm source}_{n}(E,X)}{\partial X} =  
\sum _{m}\int_{E_{\min}}^{E_{\max}}  h_{m}(E',X)  \label{for:hource} 
\end{equation}
\[
~~~~~~~~~~~~~ \left[ 
 \frac{W_{mn}(E',E)}{\lambda _{m}(E')}  \right. 
  +\left. \frac{d_m D_{mn}(E',E)}{E'\rho(X)}\right] dE'\,. 
\]

Source functions are used in the \seneca ~model \cite{Drescher:2002cr}
and by Dedenko et al. \cite{Ded04}. Particles are generated 
according to (\ref{for:hource}) and placed along the shower axis.

The source function contains a certain energy 
\[ 
E_{\rm tot}=\sum_n \int_{E_{\min}}^{E_{\max}} dE dX \frac{E  \partial h^{\rm
    source}_{n}(E,X)}{\partial X}  ~,
\]
which should be used to produce secondary particles. By choosing a
fraction $f$ of this energy, one determines the weight of the
secondaries. Ideally this should be equal to one, but this would still take
too much of CPU power. However, once 
determined, the weight stays constant even in the subsequent tracking
with MC method. This is an advantage over the thinning method where the
weight of particles is more difficult to control.

The choice of the energy threshold $E_{\max}$, where to switch
back from CE to MC is crucial. A lower value is desirable for
computation speed, 
whereas a higher value might be needed to achieve  the desired
precision for the lateral distribution function. The best value to choose
depends therefore also on the observable to be computed. For example, for
longitudinal profiles the lateral spread of muons is not important,
and one can choose a lower transition energy.

\subsection{Electromagnetic showers}

The electromagnetic part of the shower can be treated in a similar
way. 
In \seneca, pre-simulated sub-showers for a 2.5~g/cm$^2$ thick layer of
air are stored in a table. The energies are discretized in ten bins
per decade $E_i=10^{i/10}$. The table is then a matrix $V_{ij}^{mn}$,
where the indices $i,j$ represent primary and secondary energies,
$m,n$ stand for the particle types, photons and electrons/positrons.
If we have $g_i^n(X)$ particles of type $n$ and energy $E_i$, the
corresponding spectrum at $X+\Delta X$ is
\[
g_i^n(X+\Delta X) = \sum_{j,m} V_{ij}^{nm} g_j^m(X) ~~.
\]
Dedenko et al., Lagutin et al. \cite{Lagutin:1999xh}, as well as \conex
~choose to implement cascade equations 
in a similar way to (\ref{for:hce}) for all electromagnetic
interactions, i.e. pair production, bremsstrahlung, etc. 
At low energies, \conex ~implements a higher effective path length, in
order to correct for the neglected transverse dimensions. In \seneca ~one
can apply a table of pre-simulated sub-showers. This way, the
longitudinal profile can be calculated  in a quick way, using the slow
MC method only for initial fluctuations.

\section{Results and Comparisons}

\subsection{Internal checks}

\begin{figure}[bt]
\includegraphics[width=\columnwidth]{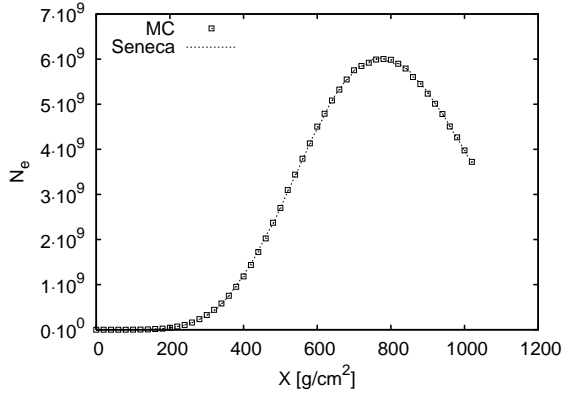}
\caption{A single mean CE shower compared with 1000 averaged
  MC-showers ($10^{-5}$ thinning).}
\label{fig:AvProfile}
\end{figure}
\begin{figure}[tb]
\includegraphics[width=\columnwidth]{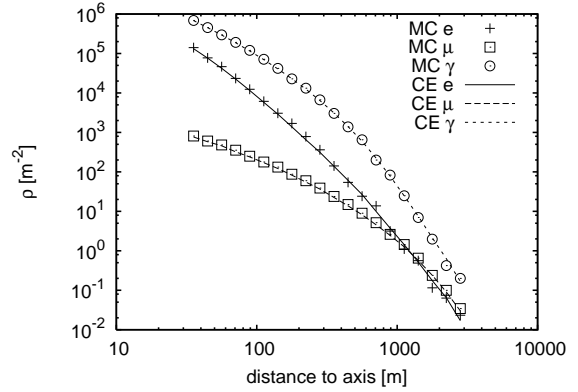}
\caption{Lateral distributions for electrons/positrons, muons and
  photons. The cut-off energies are 50~MeV for muons and 1~MeV for the
  EM-particles.}
\label{fig:AvLat}
\end{figure}

\begin{figure}[tb]
\includegraphics[width=\columnwidth]{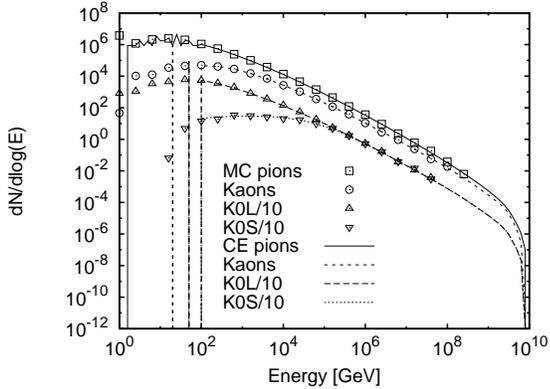}
\caption{Energy spectra for both simulation methods.}
\label{fig:Engy}
\end{figure}

\begin{figure}[bt]
\includegraphics[width=\columnwidth]{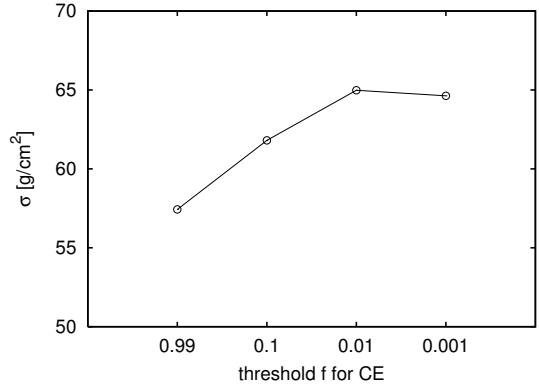}
\caption{The spread of $\Xmax$ as a function of the hybrid threshold $f$.}
\label{fig:sigmaf}
\end{figure}

\begin{figure}[bt]
\includegraphics[width=\columnwidth]{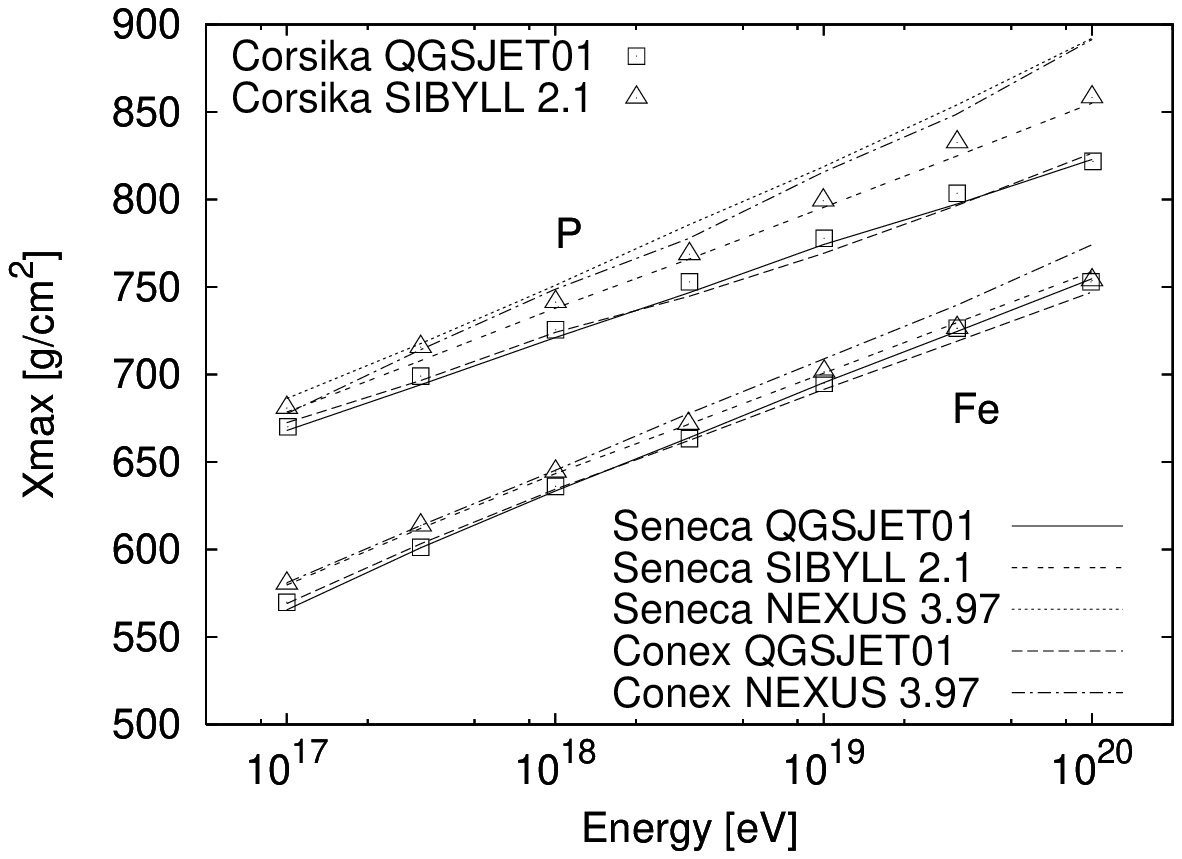}
\caption{Mean $\Xmax$ of \conex ~and \seneca ~compared with \corsika
  ~results. The Seneca proton result for \nexus ~has been computed
  with CE only   ($\Xmax$ of a mean shower).
}
\label{fig:Xmax}
\end{figure}

Precision is crucial for the hybrid approach. Any lack of precision will
appear in the solution systematically. Therefore it is necessary to
confirm whether CE and MC give the same results.
Since the cascade equations give as result a mean shower, we compare
the longitudinal profile of one CE shower to the average of 1000 MC
generated showers. Here we do not use MC for the initial part, i.e. $f=1$.
The agreement of the two curves is shown in Fig. \ref{fig:AvProfile}.

Since the CE approach is ideal to compute an average shower, one might
ask how good the maximum of a mean shower describes the mean of
fluctuating showers, as measured by experiments. 
Assuming the shower profiles are described by functions $f_i(x)$, 
the  mean shower is
\begin{equation} 
 f(x) = \frac{1}{N} \sum_{i=1}^N f_i(x) ~.
\end{equation} 
The maximum is defined by $f'(x)=0$ so $f'(\Xmax)=0$ and
$f_i'(\Xmaxi)=0$. Expanding each profile $i$ around its maximum
yields
\begin{eqnarray} 
f_i(x) &=&f_i(\Xmaxi) \nonumber  \\ 
& & +\frac{f_i''(\Xmaxi)}{2} (x-\Xmaxi)^2\nonumber
\\ & & + ... \label{for:taylor}
\end{eqnarray}
Differentiating to  find the maximum gives
\begin{eqnarray}
0 &=& f'({\Xmax}) = \frac{1}{N} \sum f_i'(\Xmax) \nonumber \\ &=& 
\frac{1}{N}\sum {f_i''(\Xmaxi)} (\Xmax-\Xmaxi) ~,
\end{eqnarray}
which leads to 
\begin{eqnarray}
\Xmax = \langle \Xmaxi \rangle ~{\rm for}~ f_i''(\Xmaxi) = c ~~. \label{for:Xmax}
\end{eqnarray}
So, if the curvature at $\Xmaxi$ is approximately a constant, the 
maximum of a mean shower corresponds to the mean of maxima. 
Simulations show that this relationship (\ref{for:Xmax}) is valid
within a few g/cm$^2$. Hence, for a quick estimate one can use a mean
shower computed with the CE approach, instead of doing 
hundreds of fluctuating showers. 
 
The same does not hold for the shower size, $\Smax$. Evaluating
formula (\ref{for:taylor}) at $\Xmax$ yields
\begin{eqnarray}
\Smax&=&f(\Xmax)=  \langle \Smaxi \rangle  \nonumber \\
&+&\frac{c}{2} \left(  \langle \Xmaxi^2 \rangle- \langle\Xmaxi \rangle^2  \right)
\end{eqnarray}
$\Smax$ of a mean shower is smaller than the mean of $\Smax$ of
many showers, since $c$ is negative. Because of energy conservation,
we expect the mean shower to be somewhat wider than on
average (larger $\Lambda$ in the Gaisser-Hillas formula).
In that sense, a mean shower is not a typical shower, but it can be
used to get a good estimate of the mean $\Xmax$. 

\subsection{Lateral Distribution Functions}

A check of lateral distribution functions (LDF) is shown in
Fig.~\ref{fig:AvLat}. The densities for electrons/positrons, muons, and
photons agree with the MC simulated spectra of a vertical $10^{19}$~eV
proton induced shower. 
In this example we compare two fluctuating showers as opposed to
comparing the average behavior. The
high energy part of the shower is calculated in the same way in both the
MC and the CE simulation. Technically this is achieved by implementing a
high energy particle stack and using the same seed for the
pseudo-random number generator.  The high energy stack takes care that
the sequence of random numbers is exactly the same until the first
particle below the threshold appears in the calculation. 
 
An important parameter for the LDFs is the threshold in energy for
 switching back to the MC method. We choose $10$~GeV for electromagnetic
 cascades and $10$~TeV for the hadronic part. This relatively high
 value is necessary to reproduce correctly the lateral spread of the
 muons. 

\subsection{Energy Spectra}

Energy spectra of all particles are directly calculated in the CE.  
An example for energy spectra at 600 g/cm$^2$ is shown in
Fig. \ref{fig:Engy}. Note
the differences in K-short and K-long below $10^5$~GeV; this is the
energy region where decays of K-shorts start dominate over interaction.

\subsection{Fluctuations}

An important check for the hybrid approach is whether it can reproduce
natural fluctuations in the $\Xmax$
distribution. Fig. \ref{fig:sigmaf} shows the spread 
$\sigma=\sqrt{ \langle \Xmax^2 \rangle-\langle\Xmax\rangle^2 }$ as a
function of the threshold $f$ (below $E=f E_0$ cascade equations are used).
As of $f=0.01$ the spread seems to converge. 
Interestingly, already $f=0.99$ reproduces 90\% of the spread,
i.e. the very first interaction is responsible for most of the
fluctuations.

\subsection{Shower maximum}

A comparison with the well tested \corsika \cite{corsika} ~model is given in
Fig. \ref{fig:Xmax}. Since the physics content in terms of external
models is the same, the hybrid approaches should give similar results. 
\qgsjet ~\cite{qgsjet}, \sibyll \cite {sibyll} and \nexus \cite{nexus}
are used as high energy hadronic models in the framework of \conex ~and
\seneca.  

  When computing an iron induced shower with CE, one does not need any
  additional tables for nucleus primaries, if the energy per nucleon
  is above the initial fluctuation threshold. 
For direct computation of a mean shower, one would have in principle to
do tables for all possible nucleus primaries up to $A=56$. In that
case, it is more reasonable to average over many iron-air collisions
(such that no nuclei are left) before solving the transport
equations.

\subsection{Arrival time}

\begin{figure}[bt]
\includegraphics[width=8cm]{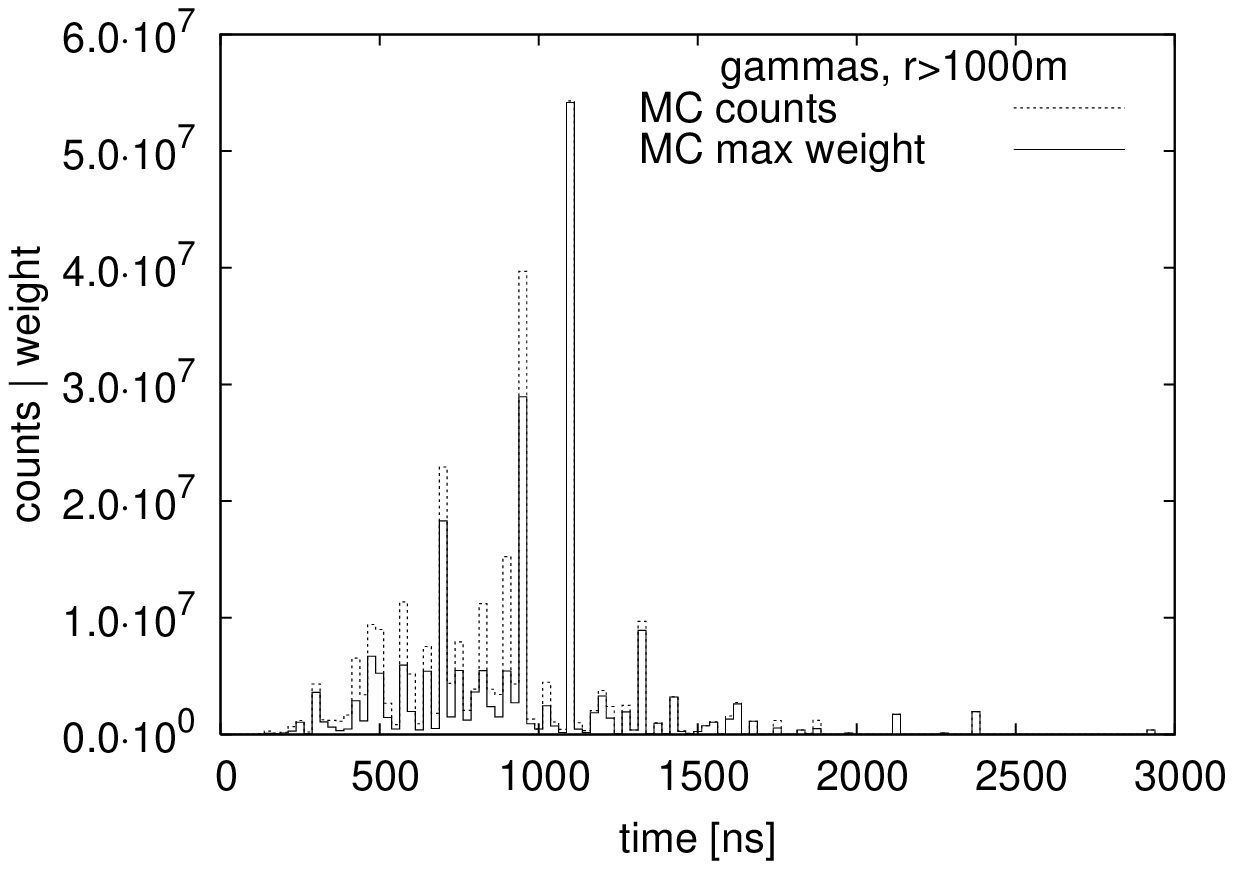}
\includegraphics[width=8cm]{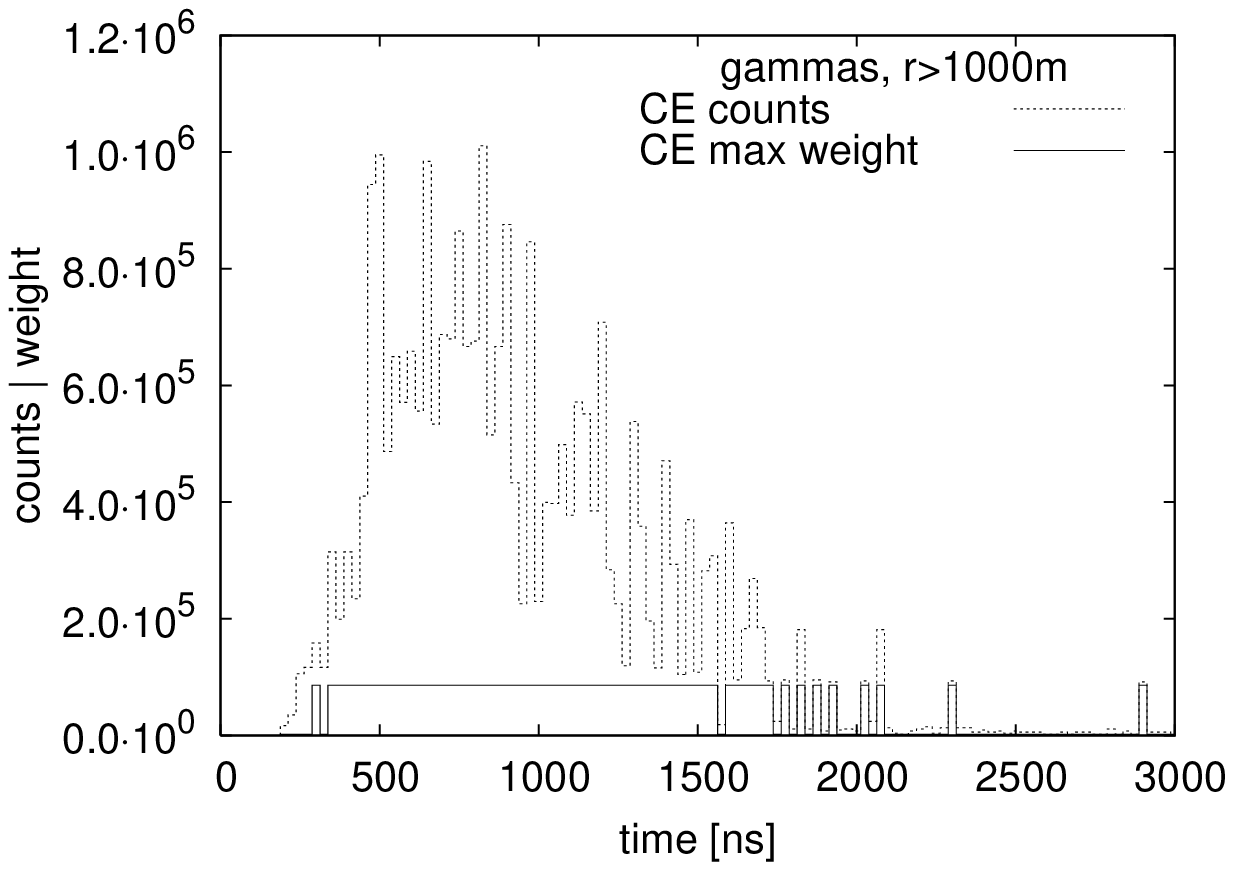}
\caption{Arrival time distributions of photons and the corresponding
  maximum weight in a given bin. The peak in the MC figure is due to a
  single particle with a huge weight.}
\label{fig:atime}
\end{figure}

Arrival times are interesting, since modern experiments try to extract
information about the primary via e.g. the rise 
time. Here, the thinning procedure has a great disadvantage, since the
weight of a given particle can be dominant for the signal, as shown in
Fig.~\ref{fig:atime}. The largest peak is due to a single particle
with a huge weight. Particles generated from the source function have
a constant weight, which is adjustable.
However, refined thinning algorithms have a weight limit, with which
the maximum weight of a particle can be defined.

This is one example where the technique of CE has a useful
side-effect, other than the main advantage of reduced CPU-time. 

\subsection{Very inclined showers}

\begin{figure}[tb]
\includegraphics[width=\columnwidth]{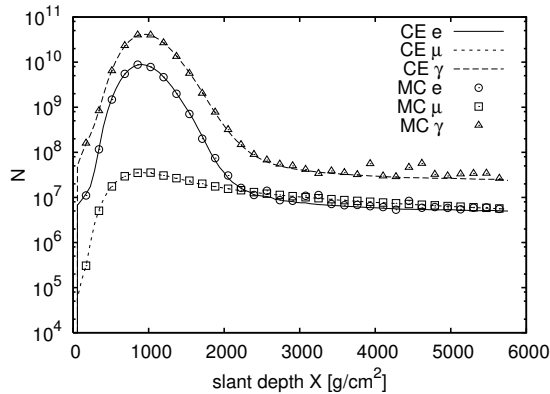}
\caption{A $80^\circ$ inclined $10^{19}$~eV proton induced shower for
  pure MC and the hybrid (CE) method.}
\label{fig:inc}
\end{figure}

\begin{figure}[bt]
\includegraphics[width=\columnwidth]{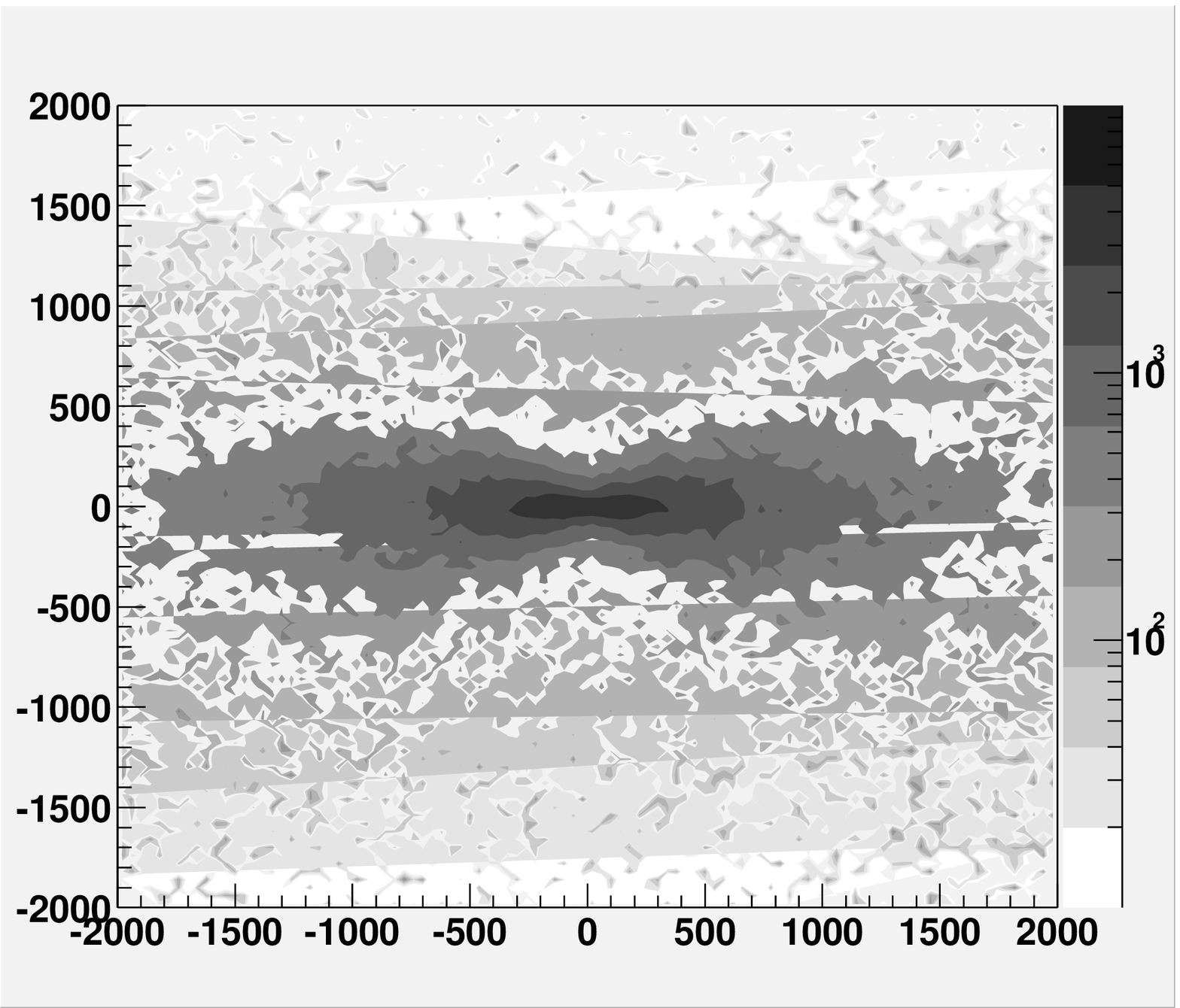}
\includegraphics[width=\columnwidth]{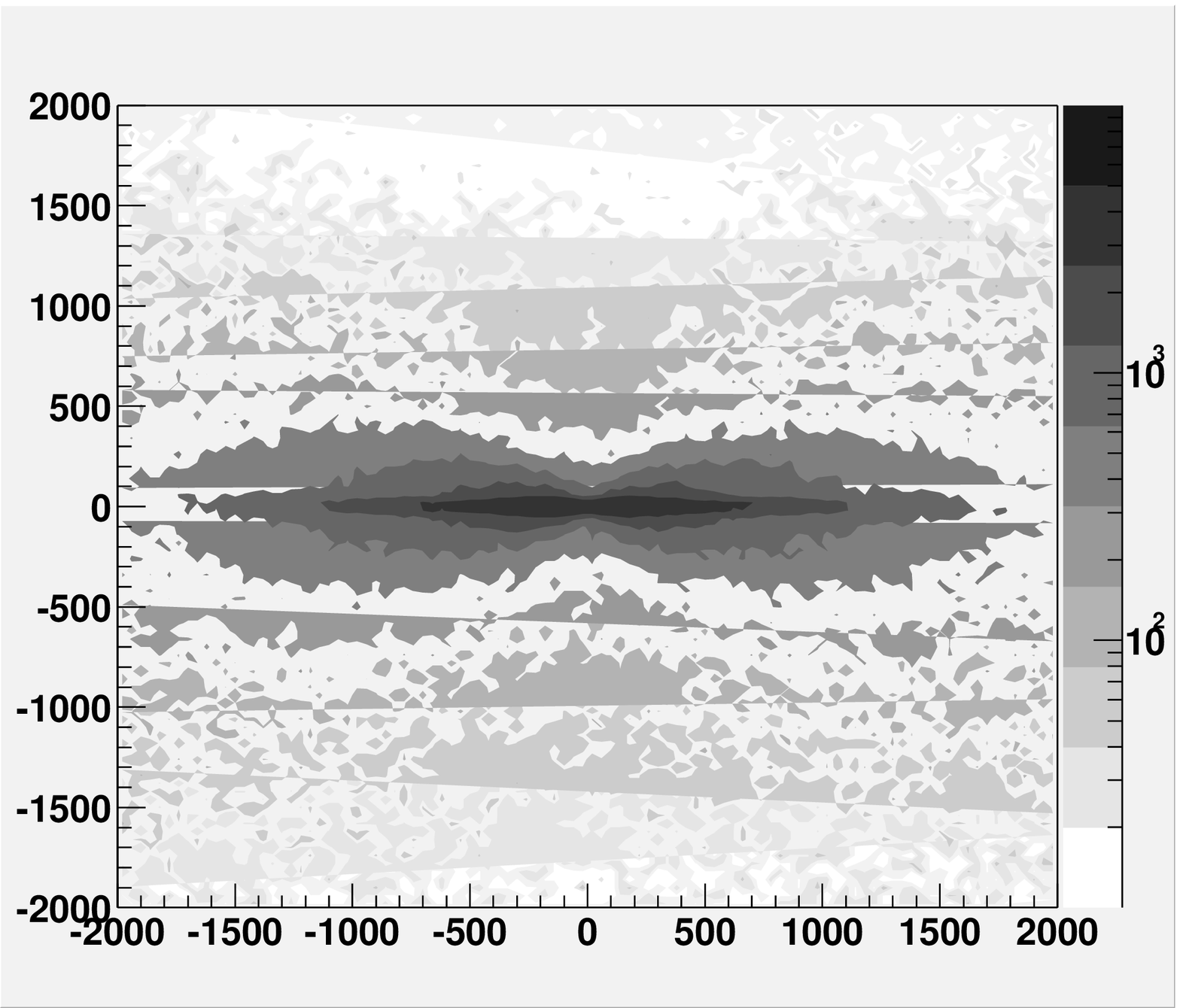}
\includegraphics[width=\columnwidth]{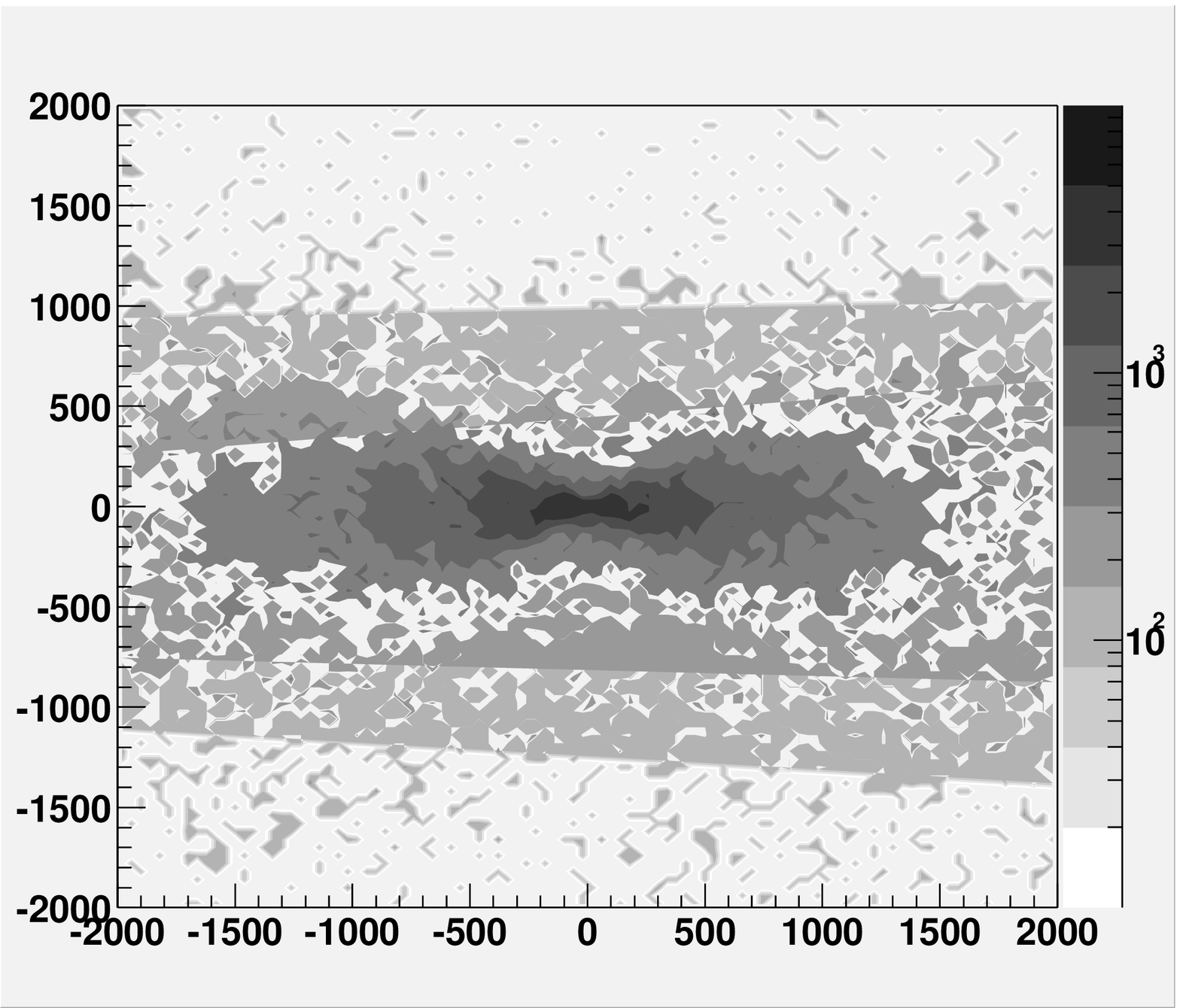}
\caption{The density of muons in the shower plane. Top panel: MC
 result, middle panel: CE result, lower panel: CE result with
 $p_t$-kick in source function.}
\label{fig:LatDens}
\end{figure}

Inclined showers are of great interest for water tank
detectors as in the  Auger observatory. These are very different from
near-vertical showers, due to a huge path-length. For 80$^\circ$ inclined
showers, we have 6 times the thickness of the atmosphere, and at
$90^\circ$ 36 times. The electromagnetic part is almost absorbed in
the atmosphere after some 2000~g/cm$^2$; only muons continue to
propagate with few interactions due to energy-loss, decays,
bremsstrahlung, etc.  The result is a relatively flat profile, see
Fig. \ref{fig:inc}. The accompanying electrons/positrons and photons
come from interactions and decays of muons. 
When calculating such a profile with CE, we solve only up to
2000~$g/cm^2$ slant depth, since beyond that value muons dominate the
shower. The profiles from the MC and CE approaches agree nicely.
But since the distance is large, a
small error in the creation of hadrons from the source function
could result in a wrong lateral shape. This is shown in
Fig. \ref{fig:LatDens}, where we plot the muon density at ground level
in the shower plane.  The deflection of positive and negative muons
due to the geomagnetic field is clearly visible. This is important to
take into account when analyzing experimental data from ground arrays. 
Fig. \ref{fig:LatDens}(top) shows the MC event and
Fig. \ref{fig:LatDens}(middle) a corresponding result with CE. Since the
particles are placed with zero angle along the shower axis, the pattern looks
distinctly different. In Fig.~\ref{fig:LatDens}(bottom) a hybrid calculation
is shown, where a mean transverse momentum is applied to all particles
generated by the source function. A value of
$p_t\approx 0.3$~GeV seems to be sufficient to reproduce the right
pattern. 

\section{CPU times}

In ref. \cite{Drescher:2002cr} it was shown that the CE approach is a
factor 20-40 faster than the MC approach, when asking for LDFs of the
same statistical quality.
A similar enhancement was found in ref. \cite{Alvarez-Muniz:2002ne} by
using the shower library approach. 

 Of course, it depends very much on the observable to be computed when
 comparing CPU times. But in general, a typical longitudinal profile
 without any lateral spread can 
 be computed in about a minute on a 1Ghz CPU. A useful lateral
 distribution function takes about 10 minutes. 

\section{Conclusions}

Hybrid calculations allow one to reduce considerably the computation
time for air shower simulations. 
The natural fluctuations arise from
the first few interactions in the atmosphere and are computed in
traditional Monte-Carlo method. Below a given threshold, particles are
then replaced with sub-showers taken from a shower library or by the
solution of cascade equations. 
The CE can be solved down to lowest energies, if 
corrections are applied to account for the neglected lateral expansion. 
The precise lateral spread can be computed again by MC method, with low
energy particles created from the source function. 

The hybrid approach, i.e. the combination of traditional Monte-Carlo
with efficient numerical methods provides a powerful tool for studying 
ultra-high energy cosmic rays. 

\section*{Acknowledgments}
The author~acknowledges support by the German Minister for
Education and Research (BMBF) under project DESY 05CT2RFA/7.
The computations were performed at the
 Frankfurt Center for Scientific Computing (CSC).

\end{document}